\documentstyle[aps]{revtex}
\input epsf         % defines \epsfbox and supporting macros
\epsfverbosetrue    % messages will show height and width

\begin{document}
%%%%%%%%%%%%%%%%%%%%%
\twocolumn[\hsize\textwidth\columnwidth\hsize\csname@twocolumnfalse%
\endcsname

\title{Temperature Effect and Fermi Surface Investigation in the Scanning
Tunneling Microscopy of Bi$_2$Sr$_2$CaCu$_2$O$_8$}

\author{K.-K. Voo$^1$, W. C. Wu$^2$, H.-Y. Chen$^3$, and C.-Y.
Mou$^{1,4}$}
\address{$^1$Department of Physics, National Tsing-Hua
University, Hsinchu 300, Taiwan\\ $^2$Department of Physics,
National Taiwan Normal University, Taipei 11650, Taiwan\\ $^3$
Texas Center for Superconductivity and Department of Physics,
University of Houston, Houston, TX 77204, USA \\ $^4$ Physics
Division, National Center for Theoretical Sciences, P.O. Box
2-131, Hsinchu 300, Taiwan}

\date{\today}
%\date{November 30, 2002}
\maketitle
\draft

\begin{abstract}
Based on a Fermi liquid picture, the temperature effect on the
impurity-induced spatial modulation of local density of states
(LDOS) is investigated for the $d$-wave superconductor
Bi$_2$Sr$_2$CaCu$_2$O$_8$, in the context of scanning tunneling
microscopy (STM). It is found that stripe-like structure exists
even in the normal state due to a local-nesting mechanism, which
is different from the octet scattering mechanism proposed by
McElroy $et~al$. {[Nature {\bf 422}, 592 (2003)]} in the $d$-wave
superconducting ($d$SC) state. The normal-state spectra, when
Fourier-transformed into the reciprocal space, can reveal the
information of the entire Fermi surface at a single measuring
bias, in contrast to the point-wise tracing proposed by McElroy
$et~al$. This may serve as another way to check the reality of
Landau quasiparticles in the normal state. We have also re-visited
the spectra in the $d$SC state and pointed out that, due to the
Umklapp symmetry of the lattice, there should exist additional
peaks in the reciprocal space, but experimentally yet to be found.
\end{abstract}

\vspace{0.1cm}
\pacs{PACS numbers: 74.70.Pq, 74.20.Rp, 74.25.Ld}
]
%%%%%%%%%%%%%%%%%%%%%%%%%%%%%%%%%%%%%%%%%%%%%%%%%%%
\section{Introduction}
\label{sec1}

A fundamental question on the high-$T_c$ cuprates remains after
more than fifteen years of the discovery of the material. It is
still not clear whether the cuprates are systems of Fermi liquid
(FL), non-FL with exotic orders such as the stripes
\cite{kivelson02}, or systems with more intricate co-existence of
different states of matter \cite{sachdev03}. In fact, this
question arises in both their superconducting and normal states.
Besides, it is not unusual that different experimental probes give
different implications -- the experimental findings are not yet
converged.

A substantial progress in the STM measurement has made it another route to
this problem. Not only
STM looks directly into the real space, but also it can be readily
connected to the reciprocal space, so-called the Fourier
transformed STM (FT-STM). Data of the low (and fixed) temperature
STM on Bi$_2$Sr$_2$CaCu$_2$O$_8$ (BSCCO) was claimed to be an
excellent manifestation of the FL behavior \cite{wang03,zhang03}.
But, since the cuprates are such involved systems, one should be
more careful to nail down the conclusion. Whether the observed STM
modulation \cite{mcelroy03} is solely the Friedel stripe arising
from the quasiparticle (QP) interference, or the Zaanen-Kivelson
stripe \cite{kivelson02} coexisting with the Friedel stripe, is
actually an issue still in debate (see the contradictory
data of Refs.~\cite{howald02} and \cite{hoffman02-2})
\cite{voo10}. Even if the stripes can be attributed to the QP
interference alone, it is still crucial to ask how well do the
quasiparticles behave? How is the extension of the ``Fermi-arc'',
and up to what temperatures they survive \cite{zaanen03}? So far
there has been few finite-temperature and normal-state STM studies
on the cuprates \cite{franz03}. These are the major concerns of
the present paper.

Based on the FL scenario, an abstract model named the ``octet''
scattering model \cite{mcelroy03} has successfully ascribed the
experimentally observed FT-STM peaks of the LDOS modulation to the
quantum interference of the QPs. Later concrete single-impurity
scattering calculations \cite{tang02,wang03,zhang03} also
supported that. The occurrence of FT-STM peaks and their evolution
with the bias change give information of the Fermi surface (FS) of
the measured system which are consistent with previous results
from the angle-resolved photoemission spectroscopy. This provides
a strong identification of a Friedel stripe out of the
Zaanen-Kivelson stripe. But nevertheless, some weak non-dispersive
peaks, which could be due to a coexisting Zaanen-Kivelson stripe,
may also exist \cite{howald02,hoffman02-2,voo10}.

It is important to investigate how far the FL picture can be
pushed, especially when the temperature is raised and the system
enters the normal state. We thus have performed similar FL-based
calculations at different temperatures to study the Friedel
stripes, for the reference of future experiments in checking the
validity of the FL picture. It is found that stripe-like structure
exists even in the normal state. This may be counterintuitive to
the octet model, since the octets should vanish in the normal
state. We argue that apart from the octets, a local-nesting
property of the FS (especially important in the normal state) can
also give rise to sizable joint-DOS, and hence sizable QP
scattering.

In Sec.~\ref{sec2}, we formulate the problem as the scattering
from a single impurity. In Sec.~\ref{secresult}, we present the
calculated spectra at different temperatures and the essential
features are remarked. In Sec.~\ref{sec3}, the physical origin of
the normal-state spectra are illucidated. It is noted that in the
normal state, the information of the entire FS can be revealed
from the data of a single measuring bias. This is in contrast to
the point-wise tracing out in the $d$SC state. Sec.~\ref{sec5} is
a closing section. In Appendix, we make a comment to the
present FT-STM data, pointing out that some additional peaks
should exist in the measurement.

%%%%%%%%%%%%%%%%%%%%%%%%%%%%%%%%%%%%%%%%%%%%%%%%%%%%%%%
\section{Formalism}
\label{sec2}

Since the single-impurity scattering model was proved to be an
excellent start for understanding the STM features in the $d$SC
state, we proceed to study the finite-temperature phenomena based
on it. We will be interested in those regions away from the
impurity neighborhood. We consider the following Hamiltonian

\begin{eqnarray} H=H_{0}+H_{\rm I}, \label{eq:htot}
\end{eqnarray}
where $H_0$ is the usual BCS mean-field Hamiltonian

\begin{eqnarray}
H_{0}=\sum_{{\bf k},\sigma}\xi_{\bf k} c^\dagger_{{\bf k}
\sigma}c_{{\bf k}\sigma}+\sum_{\bf k}\left[\Delta_{\bf k
}c^\dagger_{{\bf k}\uparrow}c^\dagger_{-{\bf k} \downarrow}+{\rm
H.c.}\right] \label{eq:hbcs}
\end{eqnarray}
with $\xi_{\bf k}$ and $\Delta_{\bf k}$ the band dispersion
relative to chemical potential and the superconducting gap
function respectively, and $H_{\rm I}$ is the part associated with
an impurity at site {\bf 0},

\begin{eqnarray}
H_{\rm I}=&&\sum_{\langle i,j\rangle,\sigma} \delta t_{ij}
c^\dagger _{i\sigma}c_{j\sigma}+\sum_{\langle
i,j\rangle}\left[\delta\Delta_{ij}c^\dagger
_{i\uparrow}c^\dagger_{j\downarrow}+{\rm H.c.}\right]\nonumber\\
&+&V_0 \left(c^\dagger_{0\uparrow}c_{0\uparrow}+c^\dagger
_{0\downarrow}c_{0\downarrow}\right).
 \label{eq:hi}
\end{eqnarray}
Parameter $\delta t$ is the deviation of local hopping,
$\delta\Delta$ is the deviation of local pairing potential, and
$V_0$ is the on-site impurity potential. This is a simplified
model but nevertheless should be enough to bring out the essential
features of interest.

It is convenient to apply the Nambu representation for the
reciprocal and real-space operators:

\begin{eqnarray}
\hat{C}_{\bf k}\equiv \left[\matrix{c_{{\bf k}\uparrow}\cr
c^\dagger_{-{\bf k}\downarrow} }\right]~~~~~{\rm and}~~~~~~
\hat{C}_{j}\equiv \left[\matrix{c_{j\uparrow}\cr
c^\dagger_{j\downarrow} }\right], \label{eq:nambu}
\end{eqnarray}
which are related to each other via the Fourier transformation

\begin{eqnarray}
\hat{C}_{\bf k}={1\over \sqrt{N}}\sum_j e^{-i{\bf k}\cdot {\bf
r}_j} \hat{C}_{j}. \label{eq:nambu.ft}
\end{eqnarray}
Here the sum is over all the lattice sites ${\bf r}_j$. Define a
$2\times 2$ energy matrix

\begin{eqnarray}
\hat{\varepsilon}_{\bf k}\equiv \left[\matrix{\xi_{\bf k}&
\Delta_{\bf k}\cr \Delta_{\bf k}^* &-\xi_{\bf k}}\right],
\label{eq:energy-matrix}
\end{eqnarray}
thus

\begin{eqnarray}
H_{0}=\sum_{\bf k}\hat{C}^\dagger_{\bf k}\hat{\varepsilon}_{\bf
k}\hat{C}_{\bf k}. \label{eq:hbcs-matrix}
\end{eqnarray}
Similarly

\begin{eqnarray}
H_{\rm I}\equiv
\sum_{i,j}\hat{C}^\dagger_{i}\hat{u}_{ij}\hat{C}_{j} + {\rm H.c},
\label{eq:hi-matrix}
\end{eqnarray}
where the matrix elements $\hat{u}_{ij}$ are to be given later.

Since we are interested in the real-space STM spectra which
measure the LDOS, we need to know the real-space, equal-site,
single-particle Green's functions. Define a single-particle
Green's function matrix

\begin{eqnarray}
\hat{G}({\bf r}_i,{\bf r}_j,\tau)=-\langle T_\tau
\hat{C}_{i}(\tau)\hat{C}^\dagger_j(0)\rangle,
\label{eq:g-matrix-r}
\end{eqnarray}
the LDOS is then given by

\begin{eqnarray}
D({\bf r},\omega)=&-&{1\over 2\pi}{\rm Im}[G_{11}({\bf r},{\bf
r},i\omega_n\rightarrow\omega+i0^+)\nonumber\\ &-&G_{22}({\bf
r},{\bf r},-i\omega_n\rightarrow -\omega-i0^+)], \label{eq:dos}
\end{eqnarray}
where $G_{\alpha\beta}$ is an element of the 2$\times$2 Green's
function matrix in (\ref{eq:g-matrix-r}), being
Fourier-transformed to the Matsubara-frequency space

\begin{eqnarray}
\hat{G}({\bf r}_i,{\bf r}_j,i\omega_n)=\int_0^\beta d\tau
e^{i\omega_n\tau}\hat{G}({\bf r}_i,{\bf r}_j,\tau).
\label{eq:g-ft}
\end{eqnarray}
Following the standard technique, the full $\hat{G}$ in
(\ref{eq:g-matrix-r}) can be expanded in terms of $H_{\rm I}$
given by (\ref{eq:hi-matrix}),

\begin{eqnarray}
&&\hat{G}({\bf r}_i,{\bf r}_j,\tau)=\hat{G}^0({\bf r}_i,{\bf
r}_j,\tau)\nonumber\\ &&+\int_0^\beta d\tau_1\langle T_\tau
\hat{C}_i(\tau)\sum_{k,\ell}\hat{C}^\dagger_{k}(\tau_1)\hat{u}_{k\ell}\hat{C}_{\ell}
(\tau_1)\hat{C}^\dagger_j(0)\rangle\nonumber\\ &&+O(\hat{u}^2)~,
\label{eq:g-matrix-full-r}
\end{eqnarray}
where, $\hat{G}^0({\bf r}_i,{\bf r}_j,\tau)\equiv\hat{G}^0({\bf
r}_i-{\bf r}_j,\tau)$ is the mean-field ``non-interacting''
$\hat{G}$ when $H_{\rm I}=0$.

The terms included in $H_{\rm I}$ in (\ref{eq:hi}) are given explicitly
here. For BSCCO, we consider a square lattice of lattice constant
$a$. In the case of a single, extended and weak impurity, we
consider local deviations $\delta t_1$ and $\delta\Delta_1$ which
couple the impurity site and its nearest neighbors, and $\delta
t_2$ and $\delta\Delta_2$ which couple the impurity's nearest
neighbors and its next nearest neighbors. Consequently, there are
17 non-vanishing $\hat{u}$ matrices in (\ref{eq:hi-matrix}):

\begin{eqnarray}
\hat{u}_{0,0}&=& \left[\matrix{V_0&0\cr 0
&-V_0}\right],~~\hat{u}_{0,\pm a\hat{x}}= \left[\matrix{\delta t_1
&\delta\Delta_1\cr -\delta\Delta_1 &-\delta
t_1}\right],\nonumber\\ \hat{u}_{0,\pm a\hat{y}}&=&
\left[\matrix{\delta t_1 &-\delta\Delta_1\cr \delta\Delta_1
&-\delta t_1}\right],\nonumber\\
\hat{u}_{a\hat{x},2a\hat{x}}&=&\hat{u}_{-a\hat{x},-2a\hat{x}}
=\hat{u}_{a\hat{y},2a\hat{y}}=\hat{u}_{-a\hat{y},-2a\hat{y}}\nonumber\\
&=&\left[\matrix{\delta t_2 &\delta\Delta_2\cr -\delta\Delta_2
&-\delta t_2}\right],\nonumber\\ \hat{u}_{a\hat{x},a\hat{x}\pm
a\hat{y} }&=&\hat{u}_{-a\hat{x},-a\hat{x}\pm a\hat{y}}
=\hat{u}_{a\hat{y},a\hat{y}\pm a\hat{x}
}=\hat{u}_{-a\hat{y},-a\hat{y}\pm a\hat{x}}\nonumber\\
&=&\left[\matrix{\delta t_2 &-\delta\Delta_2\cr \delta\Delta_2
&-\delta t_2}\right]. \label{eq:u-matrix}
\end{eqnarray}
Hamiltonians similar to this have been successfully used by Tang
and Flatt\'{e} \cite{tang02} to explain the resonant STM spectra
for Ni doped BSCCO, and by Wang and Lee \cite{wang03} and Zhang
and Ting \cite{zhang03} to explain the energy-dependent modulation
of the FT-STM spectra on superconducting BSCCO. Since we consider
the weak impurity scattering limit, $\hat{G}$ in
(\ref{eq:g-matrix-full-r}) is only calculated up to the first
order of $\hat{u}$ (the Born limit). The first order term has
already included the essential interference effect of the QPs. A
strong impurity is expected to give new features (such as a bound
state) only at the immediate neighborhood of the impurity.
Eq.~(\ref{eq:g-matrix-full-r}) is then reduced to

\begin{eqnarray}
&&\hat{G}({\bf r}_i,{\bf r}_j,i\omega_n) = \hat{G}^0({\bf
r}_i-{\bf r}_j,i\omega_n)\nonumber\\
&~~~~~~&+\sum_{k,\ell}\hat{G}^0({\bf r}_i-{\bf
r}_k,i\omega_n)\hat{u}_{k\ell} \hat{G}^0({\bf r}_\ell-{\bf
r}_j,i\omega_n). \label{eq:ladder}
\end{eqnarray}
The first term on the right is translationally invariant. Spatial
variation of the LDOS at a fixed energy comes only from the second
term. It is related to the STM spectrum discussed in
Ref.~\cite{mcelroy03} via ${\bf r}_i={\bf r}_j={\bf r}$ and
$i\omega_n \rightarrow eV + i0^+$, where $e=|e|$ is the electron
charge and $V$ is the bias voltage. Its Fourier transform is the
FT-STM spectrum. We will be discussing the second term on the
right throughout this paper.

Note that the thermal Fermi distribution function never appears in
the full Green's function $\hat{G}$ here. It arises only when the
Matsubara-frequency sum is involved, i.e., when the impurity is
dynamic and inelastic scattering occurs. Since we have assumed
an elastic impurity, our Green's function depends on temperature
only through the gap magnitude $\Delta(T)$ which should change
with temperature.

In our calculation, we have used a $800 \times 800$ square lattice
with the impurity at the center. We have chosen a simple but
reasonable impurity potential, $2\delta t_1 = 4\delta
t_2=-2\delta\Delta_1=-4\delta\Delta_2=V_0$ and have assumed that these
scales are small and in the perturbative limit. For $\xi_{\bf k}$,
we use a tight-binding band, $\xi_{\bf k}=t_1({\rm cos}k_x+{\rm
cos}k_y)/2+t_2{\rm cos}k_x{\rm cos}k_y+t_3({\rm cos}2k_x+{\rm
cos}2k_y)/2 +t_4({\rm cos}2k_x{\rm cos}k_y+{\rm cos}k_x{\rm
cos}2k_y)/2+t_5{\rm cos}2k_x{\rm cos}2k_y-\mu$ (lattice constant
$a\equiv 1$), with $t_{1-5}=-0.60, 0.16, -0.05, -0.11, 0.05$ eV
and chemical potential $\mu =-0.12$ eV, appropriate for an
optimally-doped BSCCO \cite{norman94}. In addition, the
superconducting gap is taken to be $\Delta_{\bf k}=\Delta(T)(\cos
k_x-\cos k_y)/2$ with $\Delta(T)$ the temperature-dependent gap
magnitude. Besides, we have introduced a finite broadening
$\gamma=2$ meV to the Green's function, such that $eV +i0^+$ is
replaced by $eV+i\gamma$.

%%%%%%%%%%%%%%%%%%%%%%%%%%%%%%%%%%%%%%%%%%%%%%%%%%%%%%%
\section{STM and FT-STM Spectra}
\label{secresult}

%%%%%%%
%Fig. 1
%%%%%%%

In Fig.~\ref{fig1}, we present the temperature evolution of the
real-space STM spectra at two different negative bias voltages.
The case of positive bias voltage will not be discussed as they
are qualitatively the same. The gap magnitudes are taken from
$\Delta(T=0)=44$ to $\Delta(T_c)=0$ meV to simulate the transition
from the superconducting to the normal state. At a distance of
several lattice constants away from the impurity, oscillating
Friedel stripes are seen in all panels, even in the case of normal
states [$\Delta(T)=0$]. Spectra with a similar ratio of
$e|V|/\Delta(T)$ share a similar behavior, such as those in
Fig.~\ref{fig1}(b), (c) with $e|V|/\Delta(T)=25/44, 15/26 \sim
0.57$.  On the other hand, spectra at zero $\Delta(T)$ is robust
at the change of $V$ [see Fig.~\ref{fig1}(g) and (h)]. Comparing the
relative intensities of the modulations, we see that the strongest
modulations (at a fixed
bias voltage) appear at temperatures near $T_c$. One more
important feature to note is that the different ripples live in
well-separated patches of space. The ripples do not have
large-area overlaps.

The Fourier-transforms of the above spectra are given in
Fig.~\ref{fig2}. Roughly speaking, there are two regimes for the
spectra in the $\Delta(T) \neq 0$ superconducting state, as
classified by the ratio $e|V|/\Delta(T)$. When $e|V|/\Delta(T) <
1$, there are local peaky structures that have their locations
describable by the octet model [see Figs.~\ref{fig2}(a)--(c)], in
agreement with previous studies \cite{wang03,zhang03}. Previous
studies discussed the cases with different $eV$ and a fixed
$\Delta(T)\approx \Delta(0)$, while in our case $\Delta(T)$ is
varied (by changing the temperature). In comparison of
Fig.~\ref{fig2}(b) and (c) that have a similar ratio
$e|V|/\Delta(T) \sim 0.57$, one again sees that the essential
features of the spectra depend mainly on the ratio
$e|V|/\Delta(T)$, as noted before in the discussion of the
real-space spectra. In Fig.~\ref{fig3}, a closer look of the
locations of the interference peaks in Figs.~\ref{fig2}(a)-(c) is
given. When $e|V|/\Delta(T) > 1$ [see Figs.~2(d)-(f)], some extra
peaky structures beyond the description of the octet model appear
near ${\bf q}=0$. These new structures are not expected to be
observable in real BSCCO compound because they are related to the
maximum gap part of the STM spectra (see discussion in
Sec.~\ref{sec3}), which in turn are highly inhomogeneous in space.
As $\Delta(T)$ is further decreased, the strong peak at ${\bf
q}=0$ is suppressed and vanishes at entering the normal state.

In the normal state, the FT-STM spectra are reduced to some neat
ridges instead of peaks [see Figs.~\ref{fig2}(g) and (h)], and they are
rather robust against the change of bias. They
should be readily observable in practice.

As the interference peaks in the $d$SC states (in the regime of
$e|V|/\Delta(T) < 1$) were well documented in the literature
\cite{wang03,zhang03}, we will only give a supplementary comment
on it in Appendix. In Sec.~\ref{sec3} we pay special attention
to the normal-state spectra.

%%%%%%%
%Fig. 2
%%%%%%%

%%%%%%%
%Fig. 3
%%%%%%%

%%%%%%%%%%%%%%%%%%%%%%%%%%%%%%%%%%%%%%%%%%%%%%%%%%%%%%%%
\section{Origin of the Normal-state Spectra}
\label{sec3}

In this section, we show that the normal-state FT-STM spectra have
an intimate relationship with the underlying FS.

The pronounced feature of the FT-STM in the $d$SC state is that
the peaks have locations depending on the bias. As long as the
ratio $e|V|/\Delta(T)$ is small enough to stay away from the
maximum gap region, locations of the peaks are more or less as
described by the octet model \cite{wang03} -- a model that assumes
the dominant QP scattering comes between regions of high DOS on
the FS (the octets). These are regions with the smallest velocity,
which appear at the tips of the banana-shape constant energy
contour ($|\xi_{\bf k}|^2+|\Delta_{\bf k}(T)|^2={\rm constant}$).
The agreement of the low-temperature experimental observation with
the picture was claimed to be good \cite{mcelroy03}. Later
specific single-impurity scattering calculations
\cite{wang03,zhang03} (similar to the one done in Sec.~\ref{sec2})
also supported this picture. In the normal state, such large DOS
octets no longer exist. However, we point out that there exists a
different mechanism which can also cause a substantial
$joint$-DOS, and leads to distinguished structures in the {\bf q}
space.

%%%%%%%
%Fig. 4
%%%%%%%

%
The upper panel of Fig.~\ref{fig4} shows a typical normal-state
FT-STM spectrum in an extended Brillouin zone (BZ). The underlying
FS is also shown in the lower panel. Comparing the two panels, it
is readily seen that the ``ridges'' in the spectrum are of the
same shape as the FS, but having twice the size, and differently
oriented branches overlap together. The occurrence of the ridges
can be understood from the scattering wave vectors drawn in the
lower panel. Those wave vectors, which are pivoted at ${\bf
q}=(m\pi/a$,$n\pi/a), n,m \in {\rm integer}$, are special in the
sense that they joint {\em locally~parallel} segments of the FS,
i.e., they possess a weak ``local-nesting'' \cite{voo5} (such
nesting is the weakest possible type of nesting, or it is
``marginal''). As a result, stripes in the real-space are still
understood as due to the scattering on the FS.

In the normal state, there is no small and well-separated large
DOS regions, therefore the stripes in real-space are smoothly
deformed unlike those in the $d$SC state. An implication from the
above understanding to the $d$SC state is, the new structures at
higher $eV/\Delta(T)$ in Fig.\ref{fig2} (which were not seen by
the experiment) come actually from the maximum gap regions. In
either state, quasiparticles come from different directions ${\bf
v}_F$ to hit the impurity, get bounced back, and get
interfered with some transition wave vector ${\bf q}$. Since ${\bf
v}_F$ and ${\bf q}$ are not simply related, the
orientations of the interference ripples in the patches of space
are also not simply related to their orientations from the
impurity. It is also obvious that within the same patch of space
there will be no two crossing ripples. Therefore, to explain the
``checkerboard'' pattern (overlapping ripples) seen in the
experiments using this picture, one needs to consider the existence of a
{\em dilute} concentration of impurities in the system \cite{voo4,lzhu03}.

%%%%%%%%%%%%%%%%%%%%%%%%%%%%%%%%%%%%%%%%%%%%%%%%%%%%%%%%%%%%
\section{Concluding Remarks}
\label{sec5}

Our perturbative approach should not be a concern regarding the
validity of our discussion. In this paper, it is not our purpose
to investigate the local bound states at the vicinity of the
impurity. We are interested only in those regions remote from the
impurity, where the interference between scattered quasiparticles
is expected to be the dominant process. The approach has included
the interference effect lucidly.

In the literature, it was proposed that temperature could be a
detrimental factor to the quasiparticles \cite{zaanen03}. If this
is true, the normal-state stripe described in this paper is
expected to vanish. Stripes of a different origin may exist
\cite{kivelson02}, but they are likely to differ in many aspects,
such as the orientation or temperature dependence of the stripes.
In our case, the real-space normal-state stripes \cite{voo4} and
the corresponding ridges in the FT-STM spectra reflect the
information of the entire Fermi surface. If there are destructions
of quasiparticle states at some Fermi surface segments, the
corresponding segments of the ridges in the FT-STM spectrum should
also be destroyed. The presence of a {\em pseudogap} in the normal
state should also be signified as some missing segments of the
ridges. Our study interpolates the $d$SC and normal-state STM,
using an approach supported by the experiment in the $d$SC state.
The ``fingerprints'' of the Fermi liquid under the STM probe, in
the course of the superconducting-to-normal transition, are
enumerated for future experiments.

\acknowledgements

This work is supported by the National Science Council of Taiwan
under the Grant Nos. 92-2112-M-003-009 (WCW) and 92-2811-M-007-027
(KKV and CYM). HYC thanks the support of National Center for Theoretical
Sciences (Physics Division) of Taiwan during his visit.

\appendix
\section*{Umklapp symmetry}
\label{app}

In this Appendix, a brief comment is given to the experimental
data in Ref.\cite{mcelroy03}. If one considers the octet model in
the $d$SC state more carefully, it will be interesting to note the
existence of additional peaks, which are not yet realized in
current experimental data.

%%%%%%%
%Fig. 5
%%%%%%%

In Fig.~\ref{fig5}, for a certain bias voltage, we have
illustratively shown a few direct and Umklapp scattering wavevectors in
the octet model \cite{voo9}. In addition to the
scattering vectors within the first Brillouin zone (direct), there
are also vectors connecting octets in different Brillouin zones
(Umklapp). When all these vectors are taken into account, there
should exist FT-STM peaks as shown in the lower panel of the
figure. Take an example, a new peak at the vector $q_5^\prime$ is
seen along the $q_y$ axis (when the bias voltage exceeds some
value). Peak at $q_5$ was reported by Ref.\cite{mcelroy03}, but
the corresponding $q_5^\prime$ peak was not. The more disturbing
matter is, there also exists a $q_4^\prime$ peak which may come
close to $q_{2,6}$ peaks (or vice versa speaking).

It may seem at odds that a mathematical Fourier transform of
experimental data on a discrete lattice does not automatically
possess the Umklapp symmetry. The reason may be due to the fact
that practical STM actually probes a space {\em more~continuous}
than the underlying lattice. It scans in steps {\em smaller} than
the lattice constants, and identifies the locations of individual
atoms only later, by looking at the modulation profile. The probe
tip does not jump from one atom to another right on-top!

%\bibliography{ldos}

\begin{thebibliography}{10}
\bibitem{voo10}
We think that the possible non-dispersive FT-STM peaks reported in
Ref.~\protect\cite{hoffman02-2} might be due to the bilayer
splitting of the electronic bands in the BSCCO compounds. The
bilayer splitting [$\sim ({\rm cos}k_x - {\rm cos} k_y )^2$] is
more prominent at near the antinodes, and may give rise to
splitted FT-STM peaks, while some of them could be less dispersive
at higher biases.
\bibitem{voo9} For easy comparison, our notations are chosen in a way
corresponding to those in Ref.~\protect\cite{mcelroy03}.
\end{thebibliography}
%\bibliographystyle{prsty}

\begin{figure}[p]
\begin{center}
%\mbox{\epsfxsize=1.25\hsize{\epsfbox{fig1.eps}}} \vspace{0.1cm}
\caption{Temperature dependence of the LDOS modulation surrounding
the impurity. All panels are plotted within a square region,
$r_x,r_y = -20\sim 20a$, with the impurity at the center. The left
and right columns show the spectra at bias $V$ = -15 and -25 mV
respectively, and from top to bottom the gap magnitude $\Delta(T)$
= 44, 26, 10, and 0 meV respectively. For better visualization,
spectra in different panels are plotted in different intensity
windows from $-x$ (deepest blue) to $+x$ (deepest red) as
specified by the color scale at the bottom. The individual value
of $x$ is given at the lower-right corner of the individual panel.
Those regions of the deepest blue/red have intensities below/above
the window.} \label{fig1}
\end{center}
\end{figure}

\begin{figure}[t]
\begin{center}
%\mbox{\epsfxsize=1.28\hsize{\epsfbox{fig2.eps}}} \vspace{-1.0cm}
\caption{Temperature dependence of the FT-STM spectra
(corresponding to the real-space results in Fig.~\ref{fig1}) in
the first Brillouin zone $q_x,q_y = -\pi/a \sim \pi/a$. Left and
right columns are for $V = -15$ and -25 mV respectively. From top
to bottom, $\Delta(T)= 44$, 26, 10, and 0 meV respectively. The
spectrum in each panel is plotted in an intensity window from
$x_l$ to $x_u$. The values of $x_l$ and $x_u$ are specified at the
bottom of each panel, and the color scales are shown at the bottom
of the figure.} \label{fig2}
\end{center}
\end{figure}

\begin{figure}[t]
\begin{center}
%\mbox{\epsfxsize=1.05\hsize{\epsfbox{fig3.eps}}} \vspace{-5.50cm}
\caption{This figure shows that the positions of the interference
peaks (indicated by arrows) depend mainly on the ratio
$e|V|/\Delta(T)$. Spectra in Figs.~\ref{fig2}(a)-(c) are scanned
along ${\bf q}/a=
(0.56\pi,0.56\pi)\rightarrow(0,0)\rightarrow(0.8\pi,0)$. It is
seen that Figs.~\ref{fig2}(b) and (c), which have a similar
$e|V|/\Delta(T)$ ratio, have similar peak positions and
intensities. For easier comparison, the intensity of the spectrum
in Fig.~\ref{fig2}(a) is enlarged by a factor of 2.} \label{fig3}
\end{center}
\end{figure}

\begin{figure}[t]\begin{center}
%\mbox{ \epsfxsize=1.1\hsize{\epsfbox{fig4.eps}}} \vspace{-1.5cm}
\caption{(Upper panel) The FT-STM spectrum of a normal-state
system at a bias of -23 mV. The ridge-like structure is a typical
feature of all normal-state spectra. (Lower panel) Several of the
wavevectors on the above ridges are shown in the extended
Brillouin zones. They are wavevectors having a local-nesting
property, i.e., they joint locally parallel parts of the Fermi
surface (as illustrated by pairs of red lines).} \label{fig4}
\end{center}
\end{figure}

\begin{figure}[p]
%\vspace{1cm}
%\mbox{ \epsfxsize=0.95\hsize{\epsfbox{fig5.eps}}}
%\vspace{-0.1cm}
\caption{(a) For illustration, a few of the
scattering wave vectors in the octet model (the octets are
indicated as circles) are shown. Wavevectors discussed by McElroy
$et~al.$ \protect\cite{mcelroy03} are shown in blue, while the
Umklapp wavevectors (which were not discussed in
Ref.~\protect\cite{mcelroy03}) are in red. (b) Locations of all
the wavevectors connecting all the octets. The Umklapp wavevectors
are in red.} \label{fig5}
\end{figure}

\end{document}